# Raman Response and Transport Properties of One-Dimensional van der Waals Tellurium Nanowires

Jing-Kai Qin[1,*], Pai-Ying Liao[1,*], Mengwei Si[1], Shiyuan Gao[6], Gang Qiu[1], Jie Jian[2], Qingxiao Wang[7], Si-Qi Zhang[5], Shouyuan Huang[3], Adam Charnas[1], Yixiu Wang[4], Moon J. Kim[7], Wenzhuo Wu[4], Xianfan Xu[3], Hai-Yan Wang[2], Li Yang[6], Yoke Khin Yap[5], Peide D. Ye[1,**]

[1]School of Electrical and Computer Engineering, Purdue University, West Lafayette, IN 47907, United States

[2] School of Materials Science and Engineering, Purdue University, West Lafayette, IN 47907, United States

[3] School of Mechanical Engineering, Purdue University, West Lafayette, IN 47907, United States

[4] School of Industrial Engineering, Purdue University, West Lafayette, Indiana 47907, United States

[5] Department of Physics, Michigan Technological University, Houghton, MI 49931, United States

[6]Department of Physics and Institute of Materials Science and Engineering, Washington University in St. Louis, St. Louis, Missouri 63130, United States

[7] Department of Materials Science and Engineering, University of Texas at Dallas, Richardson, TX 75080, United States

* These authors contributed equally to this work

**Address correspondence to: yep@purdue.edu (P.D.Y.)

**Abstract**

Tellurium can form nanowires of helical atomic chains. Given their unique one-dimensional van der Waals structure, these nanowires are expected to show remarkably different physical and electronic properties than bulk tellurium. Here we show that few-chain and single-chain van der Waals tellurium nanowires can be isolated using carbon nanotube and boron nitride nanotube encapsulation. With the approach, the number of atomic chains can be controlled by the inner diameter of the nanotube. The Raman response of the structures suggests that the interaction between a single-atomic tellurium chain and a carbon nanotube is weak, and that the inter-chain interaction becomes stronger as the number of chains increases. Compared with bare tellurium nanowires on $SiO_2$, nanowires encapsulated in boron nitride nanotubes exhibit a dramatically enhanced current-carrying capacity, with a current density of $1.5 \times 10^8$ A cm$^{-2}$, which exceeds that of most semiconducting nanowires. We also use our tellurium nanowires encapsulated in boron nitride nanotubes to create field-effect transistors that have a diameter of only 2 nm.

**Introduction**

Tellurium is a p-type elemental semiconductor that has recently attracted interest because of its ability to form large-area air-stable two-dimensional films with excellent transport properties[1-3]. Bulk Te has an indirect narrow bandgap of 0.35 eV with both a high electron and hole mobility of around 1000 cm$^2$/Vs at room temperature[4,5]. The broken spatial inversion symmetry of the Te atomic structure gives rise to strong spin–orbit interaction[6]. Therefore, Weyl nodes were theoretically predicted and experimentally observed deep in the valence band of Te[7,8]. Te also exhibits a strong response in terms of photoconductivity[9,10], piezoelectricity[11] and thermoelectricity[12,13].

Te also has an anisotropic atomic structure, in which atoms are strongly bonded with two nearest neighbours forming helical chains along the [0001] direction. These helical chains are stacked to form the Te crystal through weak van der Waals interaction[14-16]. This unique one-dimensional vdW crystal structure allows Te crystals to be isolated down to single atomic-chain limit[17-19]. Similar to layered two-dimensional vdW materials, which usually show layer-dependent physical properties and a superior immunity to short channel effects[20-22], this vdW-bonded 1D Te crystal composed of single-atomic chain or few-atomic chains provides a potential scaling solution for field-effect transistor (FET) applications. The structures are also expected to exhibit intriguing physics that is significantly different from that of its bulk form.

It is, however, difficult to study the physical properties of 1D free-standing Te atomic chains, since the atomic chains tend to be curled and their large surface area makes them very sensitive to the ambient environment. A practical route to stabilizing 1D nanowires is to use encapsulation by carbon nanotubes (CNTs) or boron nitride nanotubes (BNNTs), which has been successfully demonstrated for a variety of organic molecules and metals[17-19, 23-25]. The chemical stability of CNTs or BNNTs provides an effective chemical shielding for filling materials, which should also allow structure characterization. Previously, single Te chain has been successfully obtained by single-walled carbon nanotube (SWCNT) encapsulation[17], but the chain-dependent electrical and optical properties of these NWs was not obtained. The key issue with using CNTs is that their semiconducting or metallic nature will interfere with the electrical measurement of the filled materials. In contrast, electrically insulating BNNTs can provide an ideal protecting layer for such materials.

In this Article, we report the synthesis of Te NWs down to a single atomic chain and few-chain limit by filling the cavities of CNTs and BNNTs, respectively, using a physical vapor transport (PVT) technique. By controlling the inner diameter (ID) of CNTs, few-chain and single-chain Te NWs are isolated. We find that the frequency of Raman peaks of the samples varies monotonically with the number of Te atomic chains. Due to the excellent transport properties of Te and high thermal conductivity of BNNTs, the current-carrying capacity of BNNT-shielded Te NWs exceeds that of most semiconductor nanowires, reaching up to $1.5 \times 10^8$ A/cm$^2$, which is only slightly smaller than the semiconducting SWCNTs value of $4.3 \times 10^8$ A/cm$^2$.[26,27] Due to the shielding and isolation provided by the BNNTs, Te NW at the few-chain limit could stably exist in ambient, and the short-channel Te NW FETs exhibit decent electrical performance even with a diameter of only 2 nm.

**Realization and Raman response of single-chain Te NW in CNTs**

Fig. 1a shows the atomic structure of the trigonal Te lattice. Each atom is covalently bonded with its two nearest neighbors along the *c*-axis periodically to form into a helical chain, yielding an equilateral triangle projection on the basal plane. Arrays of parallel chains are stacked together by weak vdW forces and arranged hexagonally. A variety of CNTs with different ID were taken as templates (Supplementary Figure 1), and in a given batch more than 90% of the SWCNTs are successfully filled with Te atomic chains (Supplementary Figure 2). The High-angle annular dark-field scanning TEM (HAADF-STEM) and high-resolution TEM (HRTEM) images demonstrate that a single Te atomic chain can exist in ambient using a 0.8 nm SWCNT as a container, where the individual Te atoms can also be distinctively resolved with a clear 3-fold-symmetry helical structure (Fig. 1c and 1d). The successful realization of a single atomic chain is ascribed to the unique 1D vdW nature of Te. Since a single Te atom tends to covalently bond only with its two neighbouring atoms in a chain, a single atomic chain of Te can be isolated through the spatial confinement from a SWCNT. SWCNTs with an inner diameter of 1.2 and 1.5 nm increase the number of confined Te atomic chains to 2 and 3, respectively, and a larger-diameter multi-wall CNT (MWCNT) of 2.3 nm results in a wider Te NW with ~19 Te atomic chains (Fig. 1e-1g and Supplementary Figure 3). The results suggest that by carefully selecting the inner diameter of CNTs, controlled growth of ultra-narrow Te NWs with few-number of atomic chains can be realized.

As a powerful nondestructive characterization tool, Raman spectroscopy has been widely used to investigate material properties and layer-to-layer interactions, here chain-to-chain, through lattice vibrations[28]. Consistent with theoretical analysis[15,16], three first-order Raman active modes $E_1$, $A_1$, and $E_2$ located at 92 cm$^{-1}$, 123 cm$^{-1}$ and 140 cm$^{-1}$, respectively, can be clearly detected for bulk trigonal Te (Fig. 1b). The inset images in Fig. 1b illustrate the vibration pattern of Raman modes. The $A_1$ mode is caused by the chain expansion where each atom moves in the basal plane, whereas the $E_1$ and $E_2$ modes represent the bond-bending and bond-stretching with larger admixture, respectively[16]. In 2D materials, the Raman frequency is very sensitive to the number of layers, since the vdW interaction can strongly affect the electron-phonon coupling resulting in changed phonon modes[28]. Therefore, the Raman technique can also be used to evaluate the vdW interaction in Te NW crystal and identify the number of atomic chains. For a single Te atomic chain confined in a 0.8 nm SWCNT, the $A_1$ mode exhibits a large shift towards higher frequency at 196 cm$^{-1}$, while the $E_1$ and $E_2$ modes are too weak to be detected (Supplementary Figure 5). This result is in good agreement with our theoretical calculations for the free-standing single Te chain (Supplementary Figure 6 and 7). This suggests that the vdW interaction between the single Te atomic chain and the SWCNT inner wall is very weak and the Te chain could be fully relaxed in the SWCNT and exist stably in the form of a 3-fold-symmetry helical coil as shown in Fig. 1c. Fig. 1h and 1i present the evolution of Raman peak frequency with the diameter of Te NWs. As the number of Te atomic chains increases from two toward bulk (ID of CNT increases from 1.3 nm to >10 nm), clear redshifts of the $A_1$ and $E_2$ modes can be observed. For example, the $A_1$ peak is significantly shifted from 148 cm$^{-1}$ for 1.3 nm to 123.1 cm$^{-1}$ for bulk (Supplementary Figure 4). Similar to the previously reported results on few-layer tellurene[1], the unique chiral structure of Te causes the weakening of intra-chain covalent atomic bonding and the enhancement of inter-chain vdW interactions as the number of layers increases. Therefore, Te atoms could more easily restore back in bulk form, resulting in Raman frequency reduction of the $A_1$ vibration mode. On the contrary, long-range Coulumb interactions dominate the vibration mode of individual Te atomic chains in few-chain samples and lead to the hardening of intra-chain E vibration modes[29]. A greater number of Te chains would split $E_1$ mode, taking too much inaccuracy for the definition of peak positions, thus, we only focus on the $A_1$ and $E_2$ modes in this study.

DFT calculations reveal that the electronic structure of Te NWs is highly dependent on the number of chains. According to previous study[30], lone-pair and anti-bonding orbitals lead to an indirect band gap of 0.35 eV in bulk Te. As the number of chains decreases, the indirect band gap becomes more prominent. As present in Fig. 1j-1k, the band gap is expected to increase monotonically to 0.68 eV for 3 Te chains, and eventually reach 1.51 eV in the single-chain limit due to the quantum confinement effect[31]. Our theoretical results also show that adjacent chains in Te are indeed strongly coupled with each other, with a high cleavage energy of 400 meV/atom, far exceeding that of graphene (60 meV/atom). This explains the difficulty in isolating a single atomic chain of Te by the conventional exfoliation technique.

Although few-chain Te NWs could be well protected from ambient degradation by CNT encapsulation, the dominant electrical conductivity of CNTs prevents direct electrical measurements of the shielded Te NWs. BNNTs are electrically insulating with the same hollow structure as CNTs, and are also the promising 1D nanoscale containers for guest species such as metals, metal halides and molecules[24,32,33]. Here, we demonstrate that BNNTs grown by a chemical vapor deposition (CVD) process are an ideal template to encapsulate few-chain Te NWs using the same PVT technique as with CNTs (Supplementary Figure 8 and 9). Since BNNTs are electrically insulating, this property makes it possible to perform direct electrical measurements on Te NWs down to the few-chain limit[34,35].

**Realization and Raman response of few-chain Te NWs in BNNTs**

Fig. 2a is a schematic illustration of an ultra-narrow Te NW shielded by a BNNT (Te-BNNT). Before filling, the caps of nanotubes were first removed by ammonium hydroxide etching as described in the Methods Section. Fig. 2b shows a representative TEM image of an individual Te NW inside a BNNT, where the Te-filled section exhibits a clearly different contrast in the BNNT cavity. Combined with the X-ray energy dispersive spectroscopy (EDS) elemental mapping shown in Fig. 2g, we confirm that elemental Te NWs can be successfully confined inside the cavity of a BNNT. HRTEM images show that Te NWs confined in BNNTs crystallize homogeneously in the form of a single crystal, where distinctive lattice fringes belonging to the Te NW can be clearly observed, suggesting that the Te NW is well crystallized with high quality (Fig. 2c and Supplementary Figure 10). Note that the filling length of Te NWs in BNNT usually less than 400 nm, which is much shorter than that in CNTs. This is attributed to the polarity of interatomic B-N bonding and delocalization of π electrons in BNNTs, which leads to strong interface interactions between the inner BNNT surface and the filled Te atoms, restricting high aspect ratio filling[24]. In our experiments, the narrowest diameter of Te NW achieved was 2 nm, corresponding to about 19 individual Te atomic chains inside a BNNT cavity (Fig. 2d and 2e). As was previously done for CNTs, Raman frequency shifts are used to determine the diameters of Te NWs encapsulated by the BNNTs. The $A_1$ mode of a Te-BNNT with 5 nm inner diameter exhibits a blue shift of 2.4 cm$^{-1}$ compared with bulk form (about 121.2 cm$^{-1}$), and the vibration frequency continues to shift up to 129.2 cm$^{-1}$ as the diameter of the Te NW is reduced down to 2 nm (Fig. 2f and Supplementary Figure 11). Thus, we applied Raman spectroscopy to determine the diameter of Te NWs filled in BNNTs and located the samples for device fabrication after they were transferred onto a SiO$_2$/Si substrate.

**Current-carrying capacity of few-chain Te-BNNT devices**

In order to investigate the electrical properties of few-chain Te NWs, we first fabricated two-terminal devices with an isolated Te-BNNT as the channel material on a Si substrate with 90 nm SiO$_2$ as the gate dielectric. Ni/Au (30/100nm) electrodes were patterned using electron-beam lithography (EBL) followed by electron beam evaporation of metals and lift-off process. Short-channel (100 nm) devices were fabricated to ensure the channel region was fully filled by the Te NW. Empty BNNT devices were also tested for comparison. As the I-V characteristics show in Fig. 3a, the current of a Te-BNNT device with a diameter of 10 nm has about four orders of magnitude higher current carrying capacity than that of the empty

BNNT device with the same configuration, which confirms that the Te NW is responsible for the channel conduction instead of the BNNT. The device could sustain a large $V_{ds}$, and the current does not show any saturation until device breakdown at $V_{max}$=3.6 V and $I_{max}$=1.2 µA. The diameter-dependent breakdown current of ultra-narrow Te NWs was also measured based on devices with 100 nm channel lengths. Although the encapsulated Te NWs have varying diameters, BNNTs with different diameters usually have similar thickness (> 10 nm) as observed by HRTEM images, so the breakdown current is mainly determined by the intrinsic properties of the inner Te NWs. For devices with $D^2$ of 100 nm$^2$, the average breakdown current is determined to be about 1.2 µA. The reduction of $D^2$ leads to the scaling down of current carrying capacity of Te, and smallest average current measured is 62 nA for devices filled with 2 nm Te NWs. The values of ampacity in different devices are fitted linearly with $D^2$ as plotted in Fig. 3b. The realization of electrical measurement of few-chain Te NWs is attributed the efficient dissipation of Joule heating by BNNT encapsulation[36]. *In-situ* temperature-dependent Raman spectroscopy was conducted to explore the temperature limit of this passivation, and the corresponding mapping plot is present in Fig. 3c. Te NWs shielded by BNNTs could sustain high temperatures up to 350 ºC, while the naked NWs prepared by solution process were easily damaged at 250 ºC (Supplementary Figure 12 and 16). The Raman peaks of Te tend to shift towards low frequency as temperature increases, which is related to a non-crystalized phase change induced by high temperature[37].

The BN encapsulation protects the few-chain Te NW from damage during the measurements, but it also introduces a large electrical contact resistance, resulting in reduced channel current. Considering this, an accurate and selective BN etching process was developed to directly control the thinning of the BNNT insulator layer. Fig. 3d and 3e show the I-V curves of typical devices as a function of etching time. Notably, the breakdown current is greatly enhanced up to 29 µA after 20 s of etching, almost 100 times larger than the un-etched reference devices. Total resistance of the device after 20 s of etching is calculated to be 1.6 kΩ·µm, which is one order of magnitude smaller than that of devices without etching (about 21 kΩ·µm). It is worth noting that over etching the BN layer deteriorates the performance of Te-BNNT devices (Supplementary Figure 17). This result can be ascribed to a high Schottky barrier at the Ni/Te contact interface. The value of current capacity for the Te-BNNT devices reaches up to $1.5 \times 10^8$ A cm$^{-2}$, almost two orders of magnitude larger than that in bare Te NW devices of $7.4 \times 10^6$ A cm$^{-2}$ (Supplementary Figure 14). This value exceeds most of the reported semiconductor NWs[38-45] (Fig. 3f and Supplementary Table 1), and it is also higher than those from the high-ampacity TaSe$_3$ and ZrTe$_3$ quasi-1D metallic nanowires[46,47]. Indeed, it is only slightly smaller than the semiconducting SWCNTs with similar configuration (about $3.0 \sim 8.8 \times 10^8$ A cm$^{-2}$) [26-27]. Such dramatic enhancement of ampacity can be attributed to the high mobility of Te NWs, extremely high thermal conductivity of BN (about 360 W/m·K), and low electrical and thermal resistance at the BN/Te interface[48]. The excellent current carrying capacity of Te-BNNT strongly suggests that this nanomaterial system has its potential for future electronics applications, especially high-performance FETs with ultra-short channels.

**Transport properties of Te-BNNT FETs**

The electrical characteristics of FETs based on Te NWs were also systematically studied. Note that the diameter of ultra-narrow Te NWs shielded by BNNTs is restricted to less than 10 nm due to the inner diameter limitation of BNNTs, whereas Te NWs with larger diameters can be well achieved using a substrate-free solution process (see details in the Methods Section). Thus, two types of Te NWs were systematically investigated: ultra-narrow Te NWs in BNNTs with diameters of 2-10 nm and solution-grown Te NWs with diameters of 6-40 nm. Fig. 4a shows the typical schematics of FET device based on Te-BNNT, where an $Al_2O_3$ capping layer grown by low-temperature atomic layer deposition (ALD) can dope the channel and change the FET characteristics from p-type to n-type. Fig. 4b shows a false-colored scanning electron microscope (SEM) image of some short-channel devices, where the outer diameter of the Te-BNNT was measured to be 24 nm by atomic force microscopy (AFM) (Fig. 4c). Fig. 4d and 4e present the electrical performance of a Te NW FET with a diameter of 13 nm measured at room temperature. The device exhibits the typical p-type characteristics with on/off ratio on the order of $\sim 1 \times 10^4$ at a small $V_{ds}$ of 0.05 V, with on-state drain current up to 700 mA/mm at $V_g = -40$ V normalized by the diameter. Short-channel (100 nm) devices were also fabricated for comparison (Supplementary Figure 13). The diameter dependence of the on/off ratio and maximum drain current based on long-channel devices was also systematically studied, with data elucidated from more than thirty devices (Supplementary Figure 15). Due to the enhanced gate electrostatic control in narrow NWs, the on/off ratio increases sharply from $\sim 1 \times 10^2$ to $\sim 1 \times 10^5$ as the diameter decreases from 40 to 6 nm. However, the current density exhibits the opposite trend in narrower samples since they are more susceptible to surface oxidization and defects. Indeed, FETs on ultra-narrow bare Te NWs less than 6 nm in diameter start to lose electrical conductance and gate control since the crystalline structure of Te at this scale cannot be well preserved in ambient for a long time during device fabrication and measurement.

For Te-BNNT devices with \typical metal-semiconductor (MS) or metal-insulator-semiconductor (MIS) contacts, the current of channel is mainly limited by the contact resistance, determined by two factors: Schottky barrier height/width and the tunneling resistance of the inserted ultrathin insulator[49]. Although Ni has a relatively high work function which is close to the Fermi level of unintentionally *p*-doped bulk Te, direct contact at an MS interface results in a high Schottky barrier height for ultra-narrow Te NWs due to an enhanced bandgap by the quantum confinement effect as shown in Fig. 1j and Fig. 1k[50]. A large Schottky barrier height leads to a large contact resistance at the Ni/Te contact. Few-layer BNNT in the contact region after selective dry etching could provide a perfect insulating layer for the inner Te NW from the process damage. Therefore, the thickness of BN layer needs to be optimized to realize the maximum current carrying capacity in devices. The shielding of BNNT enables Te-BNNT devices to electrically functionalize with Te NWs having diameters as small as 2 nm (Supplementary Figure 18). More interestingly, the transport characteristics of Te-BNNT FETs show n-type behaviour after $Al_2O_3$ capping, indicating that the classical Fermi-level pinning issue doesn't exist on Te-BNNTs[51]. The 2 nm device possesses a large on-state current reaching up to 50 mA/mm at $V_{ds}$=1 V (Fig. 4f). Such high current is comparable to that in monolayer $MoS_2$ transistors with the same short channel length ($I_{max}$=42 mA/mm, $V_{ds}$=0.8 V, $L_{ch}$=100 nm)[52]. Note that the output curves showing non-ideal characteristics at low $V_{ds}$ are due to the insulating BN layer sandwiched at the Ni/Te interface. Fig. 4g presents the transfer characteristics of the same device at $V_{ds}$=0.6 V and 0.8 V, showing decent on-off current ratio ($>10^2$), comparable to that of BP transistors with 100 nm channel length[53]. Low-temperature ALD $Al_2O_3$ has a large amount of positive fixed charges in the dielectric and can be used as an n-type dielectric doping layer for p-type 2D Te film[5,54]. The situation here is even more interesting: The $Al_2O_3$ capping layer serves as a low work-function material to lift up the Fermi-level in the Te NW and make it an n-type FET. The large drain current obtained on n-type Te-BNNT FET reflects the fact that electron mobility of Te is as high as hole mobility in Te[5].

Fig. 4h and Fig. 4i illustrate that Te-BNNT devices enable us to extend the electrical performance to 2 nm Te NWs compared to bare Te NW devices where the diameter is limited to 6 nm. As summarized in Fig. 4j, Te NWs exhibit excellent carrier mobility larger than ~600 $cm^2 V^{-1} s^{-1}$ with a diameter of 25 nm. The reduction of mobility in narrower samples can be ascribed to surface oxidation and defects with a significant increase in surface to volume ratio. The carrier mobility decreases with smaller diameter Te NWs and the average carrier mobility for the smallest 2 nm Te NW is ~1.85 $cm^2 V^{-1} s^{-1}$ (Supplementary Figure 19).

**Conclusion**

We have reported the isolation of 1D vdW Te NWs at the few-chain and single-chain limit using CNTs as a growth template. The lattice vibrations of the sturctures exhibit a clear dependence on the number of Te atomic chains, which can be observed by peak shifts in Raman frequency. By encapsulation within BNNTs, we were able to measure the electrical transport properties for few-chain Te NWs (down to about 19 atomic chains), and both p-type and n-type Te FETs with few atomic chains as channels were demonstrated. The current carrying ability of semiconducting Te NWs exceeds the values from conventional semiconductor NWs after BNNT encapsulation and contact engineering.

**Methods**

**Encapsulating Te NW in CNTs and BNNTs.** A variety of CNTs with different inner diameters were purchased from Sigma-Aldrich. Detailed characterization of the CNTs can be seen in the Supplementary Information. Before filling, 26 mg of CNTs were heated at 420 ºC for 1h in ambient to open the ends of tubes. BNNTs were obtained by growth vapor trapping (GVT) approach (See details in Supplementary Information). The caps of the BNNTs were removed by ultrasonicating in 10% ammonium hydroxide solution for 4h, followed by thermal treatment at 800 ºC in air for 1h. Cap-opened CNTs/BNNTs and 70 mg of Te (Sigma-Aldrich, 99.8% trace metals) were sealed under vacuum ($10^{-5}$ mbar) in a quartz tube and heated at 439 °C for 4 days. As-prepared Te-CNTs or Te-BNNTs were dispersed ultrasonically into methanol for subsequent characterization.

**Preparation of Te NW by substrate-free solution process.**

Analytical-grade $Na_2TeO_3$ (0.0015 mol, Sigma-Aldrich, 99.9% trace metals) and PVP (1g, Sigma-Aldrich, 99.9% trace metals) were dissolved in distilled water (100 ml) to form a homogeneous solution, which was sealed into a Tefon-lined stainless-steel autoclave together with an aqueous ammonia solution (25%, wt/wt%) and hydrazine hydrate (80%, wt/wt%). The reaction was conducted at 180 ºC for 3 h, and then cooled to room temperature naturally. Finally, the products were precipitated by centrifugation at 5,000 rpm for 5min and washed with distilled water to remove residuals.

**HRTEM and Raman characterization.** A dispersion of Te-CNTs or Te-BNNTs was drop cast onto a Si substrate with 90 nm $SiO_2$ as gate dielectric for Raman analysis. To eliminate the influence of CNT radial breathing modes ranging from 150 $cm^{-1}$ to 180 $cm^{-1}$, AFM was first used to locate the testing area where CNTs was sparsely distributed on substrate[55]. Raman measurements were performed using a HORIBA LabRAM HR800 Raman spectrometer. The system is equipped with a He-Ne excitation laser with 633 nm wavelength. Before measurement, the system was calibrated with the Raman peak of Si at 521 cm$^{-1}$. *In-situ* temperature-dependent Raman spectroscopy was conducted by heating samples on a hot plate from room temperature to 400 ºC in steps of 50 ºC. Spectra were only collected after the temperature remained stable for 5 min in ambient. For TEM characterization, the powder sample of CNTs was first dissolved in methanol, followed by 10 minutes of ultrasonic treatment. Then three drops of the solution were dropped onto a 3 mm TEM grid supported by lacey carbon film. High-resolution TEM (HRTEM) was performed with FEI Talos F200x equipped with a probe corrector. This microscope was operated with an acceleration voltage of 200 kV. Scanning transmission electron microscopy (STEM) imaging was carried out using a JEM-ARM200F microscope equipped with a CEOS probe aberration corrector. To reduce beam damage of the CNTs and encapsulated Te, the microscope was operated at 80 kV. The convergence semi-angle and the current of the electron probe were 25 mrad and 23 pA, respectively. High angle annular dark-field (HAADF) imaging was performed using a JEOL dark field detector, with the collection semi-angle ranging from 40-150 mrad.

**Device fabrication and electrical measurement.** After determining the diameter of Te NWs encapsulated in BNNT by Raman spectroscopy, source/drain regions were patterned by EBL, followed by electron beam evaporation of 30 nm Ni and 100 nm Au as metal contacts. The devices were measured with a probe station connected to a semiconductor characterization system (4200SCS, Keithley) at room temperature. For thinning of BN at the contact areas,

Ar/SF$_6$ plasma with power of 75 W and pressure of 50 Pa was used with different irradiation times at room temperature.

**DFT calculations.** The band structure and phonon energy were obtained from density functional theory (DFT) calculations using the Quantum ESPRESSO package. The norm-conserving pseudopotential within the generalized gradient approximation of Perdew-Burke-Ernzerhof (GGA-PBE) is used. The plane-wave basis is set to an energy cutoff of 60 Ry. The k-point sampling of the Brillouin zone is 1×1×12 for the single chain and 3-chain bundle, and 8×8×6 for the bulk tellurium. The structure was relaxed until the total energy converged to within 1×10$^{-6}$ Ry. The one-dimensional chains are separated with a vacuum region over 15 Å with its periodic images. The band structures are calculated including the spin-orbit interaction.

**Data availability.** The data that support the plots within this paper and other findings of this study are available from the corresponding author upon reasonable request.

**Acknowledgements**

The work was supported in part by U.S. Army Research Office and in part by ASCENT, one of six centers in JUMP, a Semiconductor Research Corporation (SRC) program sponsored by DARPA. J.J. and H.W. acknowledge the support from the U.S. Office of Naval Research for the TEM effort. S.G. and L.Y. are supported by the National Science Foundation (NSF) CAREER Grant No. DMR-1455346 and the Air Force Office of Scientific Research (AFOSR) grant No. FA9550-17-1-0304. M.J.K was supported in part by Global Research and Development Center Program (2018K1A4A3A01064272) and Brain Pool Program (2019H1D3A2A01061938) through the National Research Foundation of Korea (NRF) funded by the Ministry of Science and ICT. The computational resources have been provided by the Stampede of Teragrid at the Texas Advanced Computing Center (TACC) through XSEDE.

**Supplementary Information**

Additional details for Raman characterization of CNTs, HRTEM and HAADF-STEM characterazation of single atomic Te chain in CNTs, photon energy calculations, preparation and characterizaiton of high-quality BNNTs, preparation and characterizaion of solution-grown Te NWs, electrical measurements of Te NW transistors, current ampacity measurements of Te-BNNTs under over-etching condition, thermal stability of Te NWs and Te-BNNTs, comparision of ampacity between different semiconductor NWs and the evaluation of contact resistance in Te-BNNT devices are in the supplementary information.

**Author Contributions**

J.K.Q and P.Y.L contributed equally to the work. P.D.Y and J.K.Q conceived the idea and proposed the Te-CNT and Te-BNNT research. J.K.Q and P.Y.L performed the growth experiments and analyzed the experimental data. J.K.Q, P.Y.L, M.W.S, G.Q and A.C performed device fabrication and analyzed the experimental data. S.Q.Z and Y.K.Y prepared the BNNT samples. Y. W and W. W synthesized bare Te nanowires. S.Y.H and X.F.X conducted and supervised the Raman measurement. J.J, Q.W, M.J.K, and H.Y.W performed and analyzed the STEM measurement. S.Y.G and L.Y performed and supervised the DFT calculation. J.K.Q, P.Y.L, and P.D.Y co-wrote the manuscript.

**Financial Interest Statement**

The authors declare no competing financial interest.

**Figure 1**

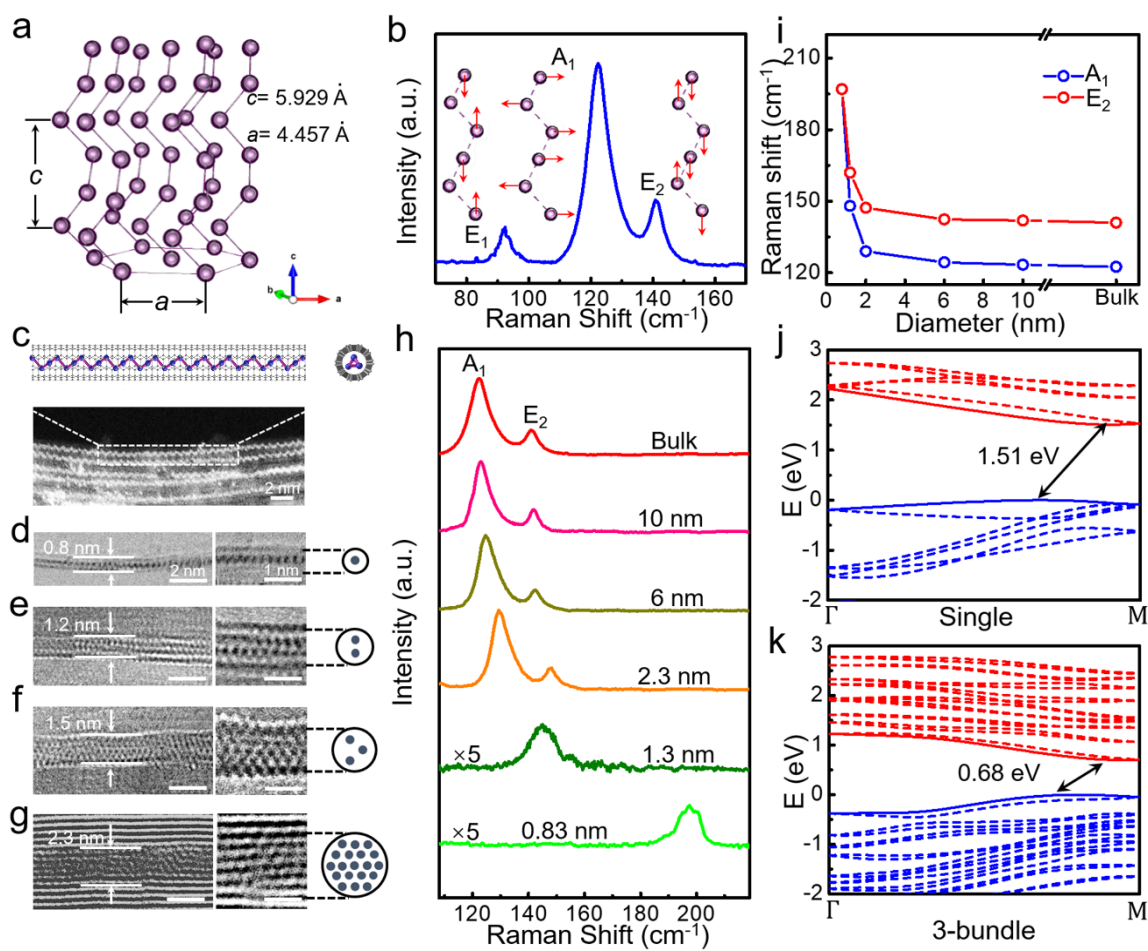

**Fig. 1 | Characterization of Te NWs isolated down to few-chain limit by CNT encapsulation. a,** Atomic structure of hexagonal Te. **b**, Raman spectrum obtained in bulk Te crystal. Inset images show the displacement pattern for vibration mode: $E_1$, $A_1$, and $E_2$. **c,** HAADF-STEM image of single Te chain in 0.8 nm SWCNT, where clear 3-fold-symmetry helical coils are distinguished. Inset image shows the corresponding schematic illustration. **d-g,** HRTEM images of the single, double, triple and 19 chains of Te shielded by CNT. **h,** Raman spectrum evolution of Te NWs with different ID of CNTs. **i,** Dependence of the shifts of Raman peak frequency on ID of CNTs. **j,** Band structure of a single Te atomic chain. **k,** Band structure of triple Te atomic chains.

**Figure 2**

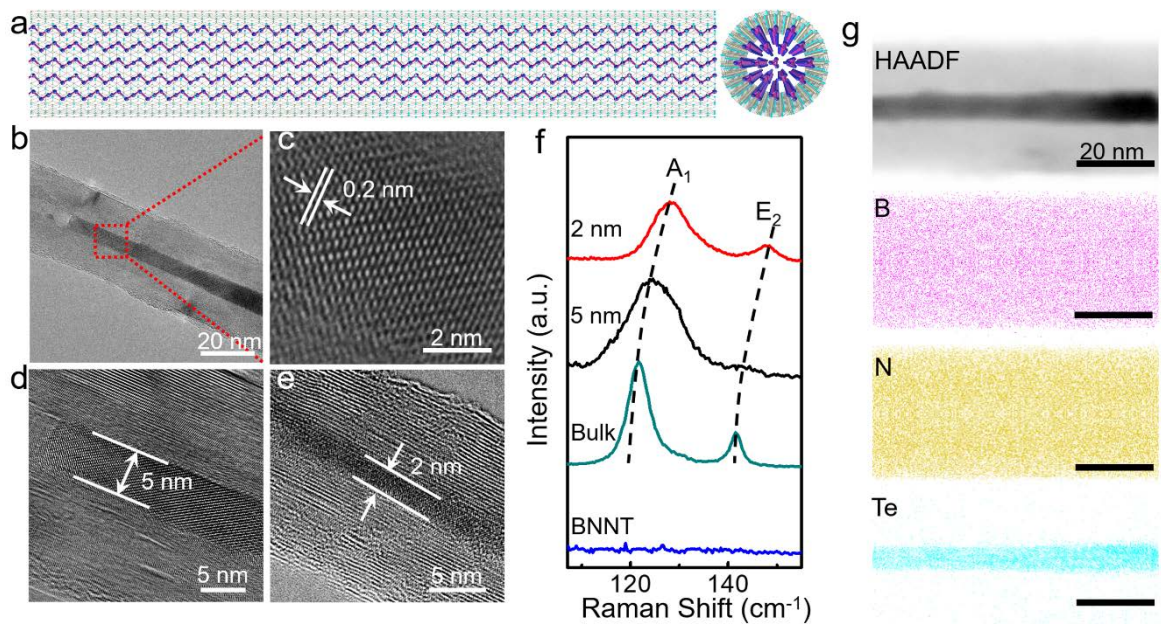

**Fig. 2 | Characterization of few-chain Te NWs shielded by BNNTs. a,** Schematic illustration of a Te-BNNT. **b,** TEM image of a 5 nm Te NWs in a BNNT. **c,** Enlarged HRTEM image of red rectangular region marked in Fig. 2b. **d-e,** HRTEM images of BNNT filled with 5 nm and 2 nm Te NWs, respectively. **f,** Raman spectrum comparison of Te NWs in BNNTs with different diameters. **g,** HAADF-STEM image and EDS maps showing the chemical composition of the filling material.

**Figure 3**

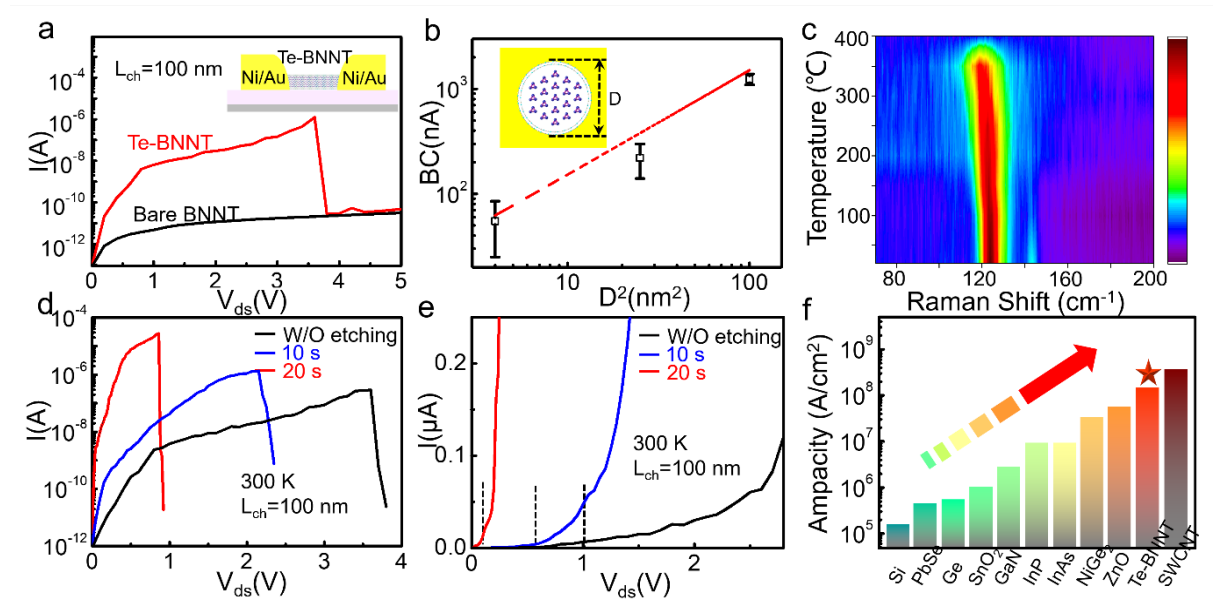

**Fig. 3 | Current-carrying capability of Te NWs encapsulated in BNNTs. a,** I-V curves up to breakdown of a Te-BNNT device with a 100 nm channel length and an empty BNNT device. **b,** Breakdown current (BC) versus $D^2$ with all channel lengths of 100 nm. **c,** Mapping plot of *In-situ* temperature-dependent Raman spectra of a Te-BNNT. **d,** Logarithmic scale and **e,** Linear scale plot of I-V characteristics in Te-BNNT devices with different etching time. **f,** Comparison of ampacity of Te-BNNTs with other semiconductor NWs.

**Figure 4**

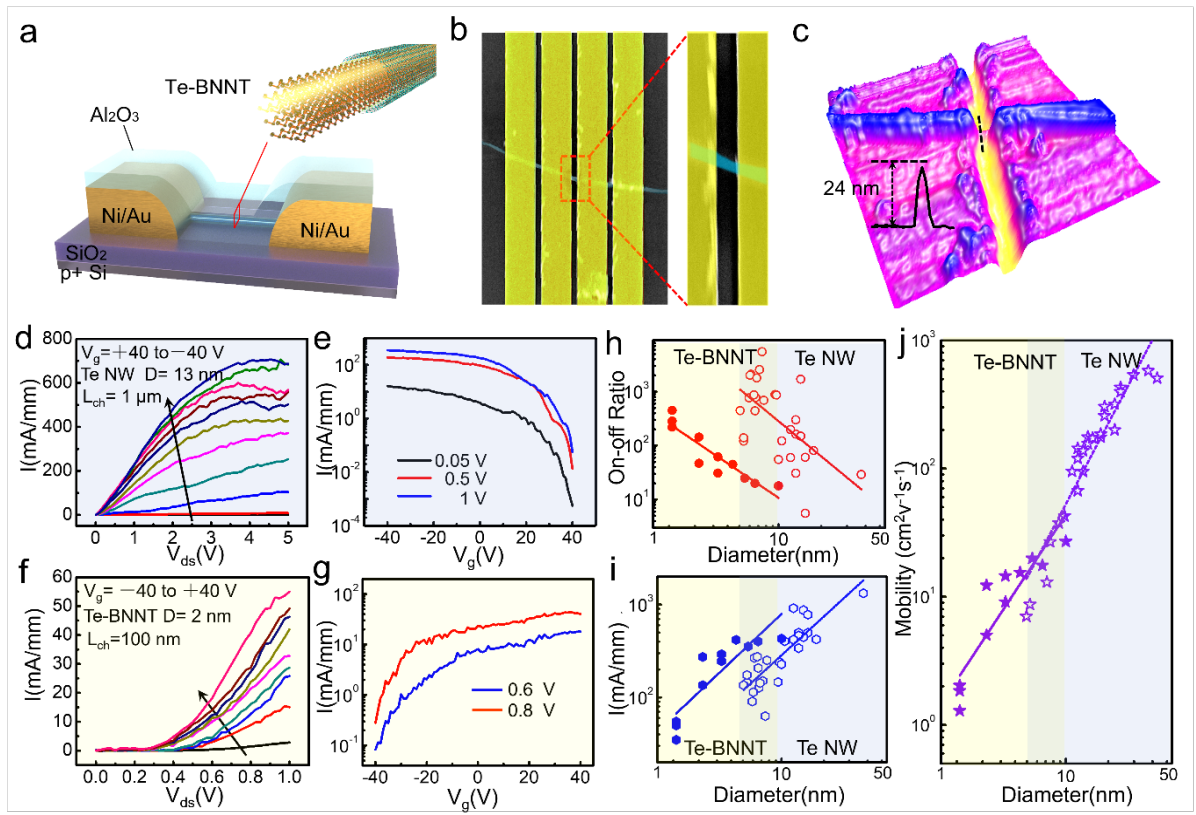

**Fig. 4 | Electrical measurements of transistors based on few-chain Te NW. a,** Schematic illustration of an individual Te-BNNT FET. **b,** False-colored SEM image and **c,** AFM height profile of a representative FET device before $Al_2O_3$ capping. **d,** Output characteristics of a typical p-type Te NW transistor with a diameter of around 13 nm. **e,** Corresponding transfer curves of the same device. **f,** Output characteristics of a n-type Te-BNNT transistor with a diameter of around 2 nm after 15 nm $Al_2O_3$ film capping. **g,** Corresponding transfer curve, exhibiting typical n-type transport behaviour with on/off ratio of 450 at $V_{ds}$=0.6 V. **h,** On-off ratio **i,** current density at $V_{ds}$=1 V and **j,** carrier mobility of Te-BNNTs and Te NWs short-channel FETs. The solid signs represent Te-BNNT NW devices and the empty signs represent bare Te NW devices.

Supplementary Information for

# Raman Response and Transport Properties of One-Dimensional van der Waals Tellurium Nanowires

Jing-Kai Qin[1,*], Pai-Ying Liao[1,*], Mengwei Si[1], Shiyuan Gao[6], Gang Qiu[1], Jie Jian[2], Qingxiao Wang[7], Si-Qi Zhang[5], Shouyuan Huang[3], Adam Charnas[1], Yixiu Wang[4], Moon J. Kim[7], Wenzhuo Wu[4], Xianfan Xu[3], Hai-Yan Wang[2], Li Yang[6], Yoke Khin Yap[5], Peide D. Ye[1,**]

[1]School of Electrical and Computer Engineering, Purdue University, West Lafayette, IN 47907, United States

[2] School of Materials Science and Engineering, Purdue University, West Lafayette, IN 47907, United States

[3] School of Mechanical Engineering, Purdue University, West Lafayette, IN 47907, United States

[4] School of Industrial Engineering, Purdue University, West Lafayette, Indiana 47907, United States

[5] Department of Physics, Michigan Technological University, Houghton, MI 49931, United States

[6]Department of Physics and Institute of Materials Science and Engineering, Washington University in St. Louis, St. Louis, Missouri 63130, United States

[7] Department of Materials Science and Engineering, University of Texas at Dallas, Richardson, TX 75080, United States

* These authors contributed equally to this work

**Address correspondence to: yep@purdue.edu (P.D.Y.)

1. **Confirmation of Te filling in single-wall carbon nanotubes (SWCNTs)**

Four types of CNTs are used as containers for Te filling, and all of them are purchased from Sigma-Aldrich. The detailed characterization is listed below:

(1) SWCNTs are produced by the CoMoCATTM catalyst chemical vapor deposition (CVD) process with diameters ranging from 0.7-1.3 nm, and the stated average diameter is to be 0.82 nm. The SWCNTs are provided in powder form with more than 99% as SWCNT.

(2) SWCNTs with average diameter of 1.3 nm, prepared by arc plasma jet method using Ni and Y as catalyst. The SWCNTs are supplied in powder form with very high purity (>99% as carbon nanotube).

(3) MWCNTs with median diameter of 2.5-3.0 nm, prepared by patented CoMoCAT synthesis process, and the CNT purity is larger than 94% measured by TGA.

(4) MWCNTs, the outer diameter is 6-10 nm and the length is 0.5-10 μm. They are prepared by arc ablation method with MWCNT basis around 30%.

All the CNTs were characterized using Raman spectroscopy as shown in **Supplementary Figure** 1.

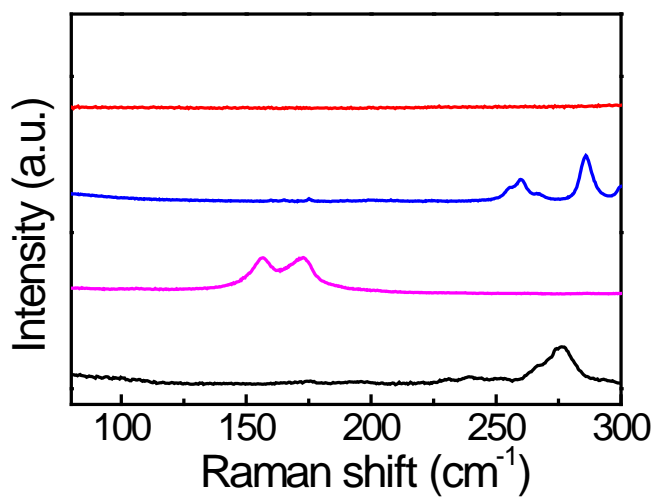

**Supplementary Figure 1 | Raman spectrum of CNTs with the laser wavelength at 633 nm.**

The SWCNT powder was first thermal treated at 500 ºC for 1h in air condition to open the tube ends, which enables the efficient filling of ultra-narrow Te NW. The as-grown Te-SWCNTs were ultrasonically dispersed in methanol solution and dropped on TEM microgrids for characterization. As the high-resolution TEM (HRTEM) image shown in **Supplementary Figure** 2, clear lattice fringe could be observed in the cavity of SWCNTs, indicating most of the SWCNTs were filled with Te NWs. The HAADF-STEM image could provide Z-contrast with atomic lateral resolution (Z =atomic number), by which we could determine the filling yield. Combined with a GIF Quantum spectrometer (Gatan) specialized for low-voltage operation and EDX mapping image of Te-SWCNTs (Inset image in **Supplementary Figure** 2b), we can determine that more than 90% of the SWCNTs were filled with Te atomic chains.

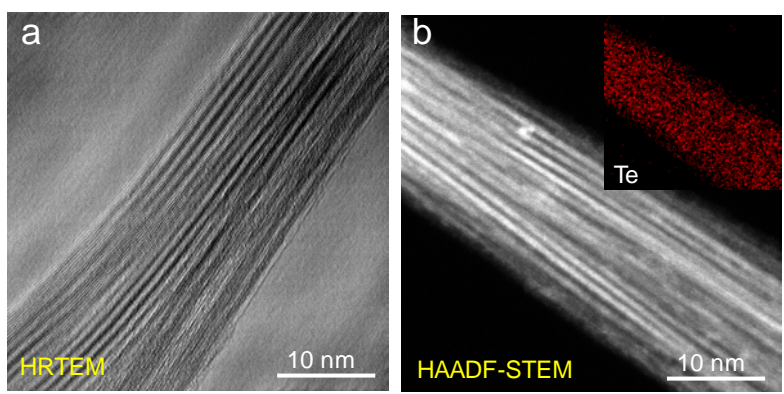

**Supplementary Figure 2 | TEM characterization of Te chains in SWCNTs**. **a,** HRTEM image. **b,** HAADF-STEM image. Inset shows the corresponding EDX mapping image.

The HAADF-STEM image could further demonstrate that the single Te atomic chain is inside the SWCNT. As shown in **Supplementary Figure** 3a, the 3-fold-symmetry helical coils of Te atoms can be clearly observed. In addition, larger Te NWs can be successfully obtained with the increase of CNT diameters (**Supplementary Figure 3b and 3c**). The results suggest that by carefully selecting the inner diameter of CNTs, controlling growth of ultra-narrow Te

NWs with few-number of atomic chains could be realized.

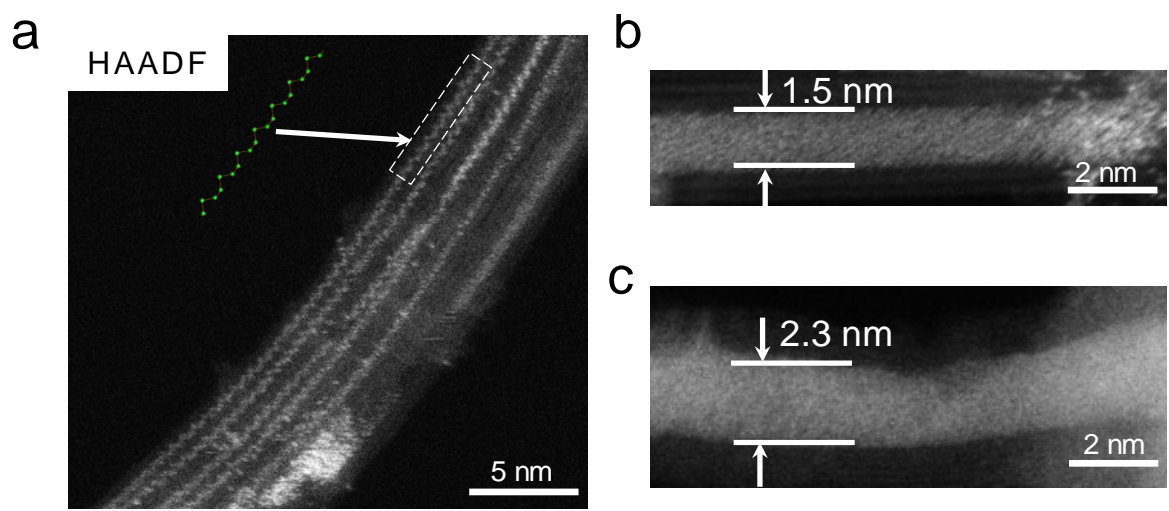

**Supplementary Figure 3 | Characterization of ultra-narrow Te NWs isolated down to few-chain limit by CNT encapsulation. a,** HAADF image of a single Te chain in 0.8 nm SWCNT, where the 3-fold-symmetry helical coils can be clearly distinguished. **b-c,** HAADF images of larger Te NWs in CNT with a diameter of 1.5 and 2.4 nm, respectively.

In the experiment, AFM was first used to locate the testing area where CNTs was sparsely distributed on the substrate. As shown in **Supplementary Figure 4a-c**, the individual SWCNTs could be clearly identified. The corresponding height profile and distribution statistics demonstrated that the outer diameter of SWCNTs ranges from 1.1-1.5 nm with the average value of 1.3 nm. In this way, the interference between SWCNTs towards the Raman vibration could be eliminated and the peak intensity of CNT radial breathing modes are greatly suppressed. **Supplementary Figure 4d** shows the corresponding Raman spectra of bare SWCNTs and Te-SWCNTs. By carefully eliminating the scattering information of SWCNTs, we can obtain the $E_1$ and $A_1$ peaks of single atomic Te chains, which locate at 83 cm$^{-1}$ and 148 cm$^{-1}$, respectively, while the $E_2$ mode exhibits extremely small intensity. Note that the relative intensity ratio ($E_1/A_1$) of single Te atomic chains increases largely up to 1.1, much larger than that in bulk form (~0.26). It means the $A_1$ mode of single Te atomic chains is significantly suppressed due to the SWCNT encapsulation.

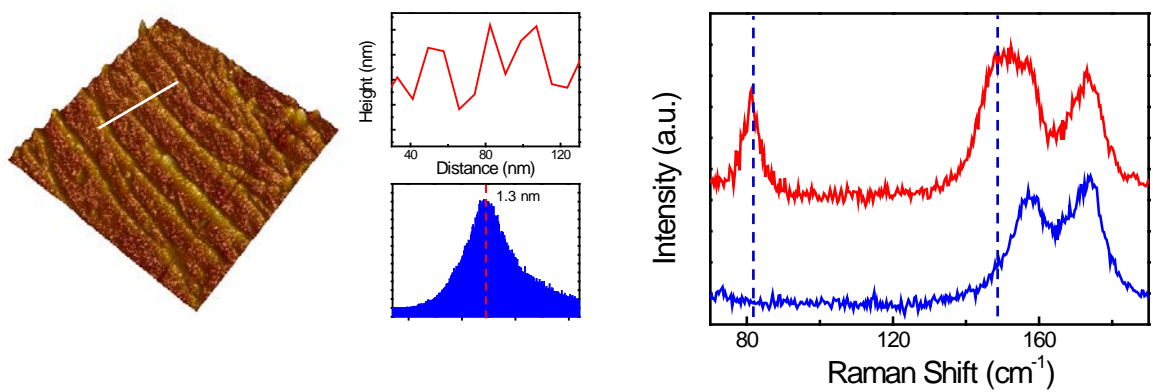

**Supplementary Figure 4 | Raman characterization of Te atomic chains in SWCNT. a,** 3D view of SWCNTs sparsely distributed on substrate. **b-c,** the corresponding height profile and distribution statistics of SWCNTs. **d,** The Raman spectra of bare SWCNTs and Te-SWCNTs.

The Raman spectrum obtained from single Te chains encapsulated in 0.8 nm SWCNTs shows only one strong peak at around 196 cm$^{-1}$, while other peaks are too weak to be detected (Supplementary Figure 5). This is understandable, since the energies of the $A_1$ and $E_2$ modes are closer, this significant change of the $A_1$ mode also impacts the energy and vibration of the $E_2$ mode and leads to the hybridization.

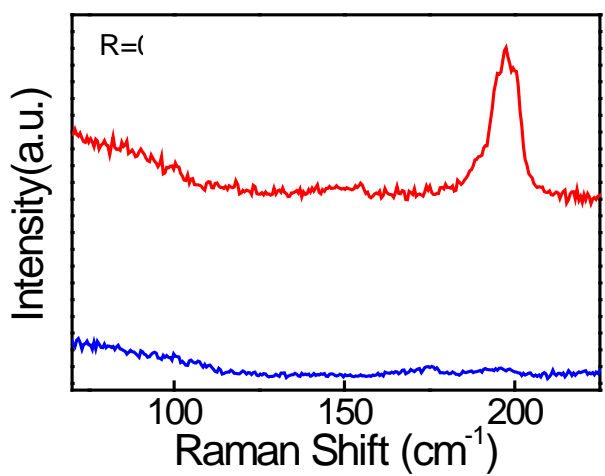

**Supplementary Figure 5 | Raman characterization of single Te atomic chains in SWCNTs.**

## 2. Photon energy calculation of Te atomic chain encapsulated in SWCNTs

As illustrated in Supplementary Figure 6a, free-standing single atomic Te chain was fully relaxed to obtain its photon energies [1-3]. DFT-relaxed structures give the frequency of $E_1$, $E_2$ and $A_1$ modes to be 68, 178 and 192 cm$^{-1}$, respectively. With the increase of chains, both the $A_1$ and $E_2$ would move toward the low-frequency direction, while the $E_1$ mode shows the opposite tendency, which are well matched with experiments. The DFT results suggest that vdW interaction between the Te chain and CNT is very weak and the CNT encapsulation keeps the free-standing structure of Te atomic chain.

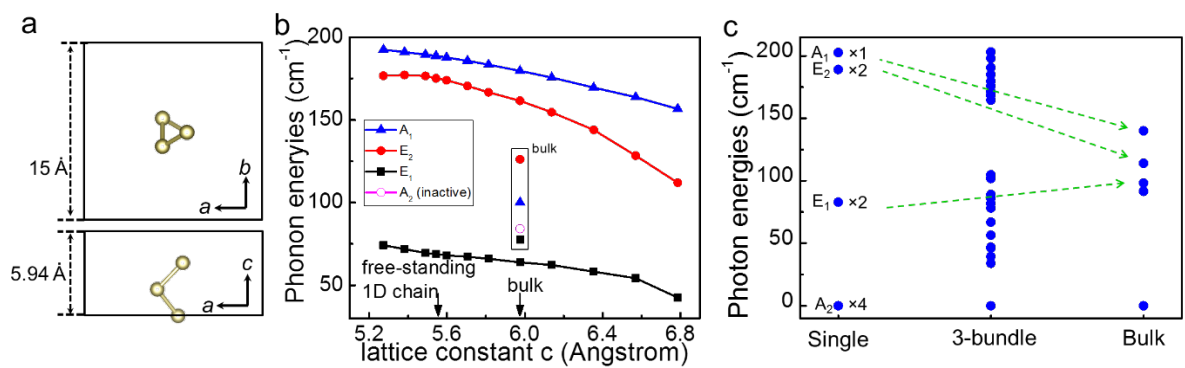

**Supplementary Figure 6 | The photon energy calculation of Te atomic chains. a,** Atomistic structure of the Te atomic chain. **b,** Phonon energies of a single atomic Te chain with dependence of lattice constant. **c,** Phonon energies for a single Te atomic chain and bulk Te. In 3-bundle case, the interaction between chains would split the Raman vibration modes and result into the increase of modes, while these modes are degenerated in single chain case or in bulk form.

It should be noted that defective 3H structure could be also energetically stable except for ideal 3H structure with SWCNT diameters larger than 1.1 nm. As shown in Supplementary Figure 7a, this screwed structure can be split into a possessing left- and another right-handed chain. The corresponding electrical band structure and photon structure are present in Supplementary Figure 7b-c and Supplementary Figure 7e-f, respectively.

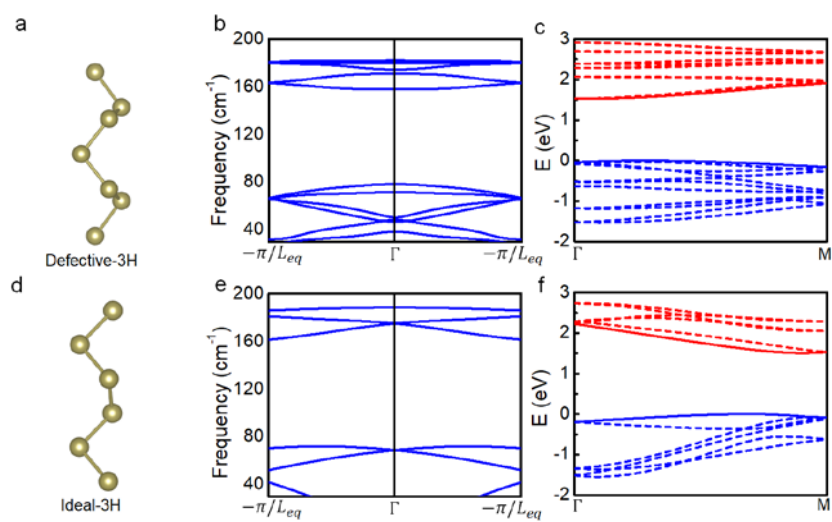

**Supplementary Figure 7 | DFT calculation of defective 3H structure and ideal 3H structure. a,** Atomistic structure of defective 3H structure. **b,** Corresponding phonon energies and **c,** electrical band structure. **d,** Atomistic structure of ideal 3H structure. **e,** Corresponding phonon energies and **f,** electrical band structure.

### 3. Ultra-narrow Te NWs shielded by BNNTs with different diameters.

BNNTs are obtained using chemical vapor deposition (CVD) combined with growth vapor trapping (GVT) approach as reported in our previous work[4] (Supplementary Figure 8). In the beginning, clean Si substrate was uniformly covered by 10 nm $MgCl_2$ films deposited by pulsed laser deposition (PLD) technique, and it was then placed upside down on the top of $Al_2O_3$ boat with B, $MgCl_2$, and FeO precursors (molar ratio of 4:1:1) and loaded at the end of a quartz tube. The quartz tube was placed in the center of a horizontal tube furnace with the holding temperature of 1200 °C for 30 min, and the ammonia flow was set to be 200 sccm. The GVT is realized since the substrate is placed at the end of quartz tube, and the growth process cannot be affected by the flow of the ammonia gas.

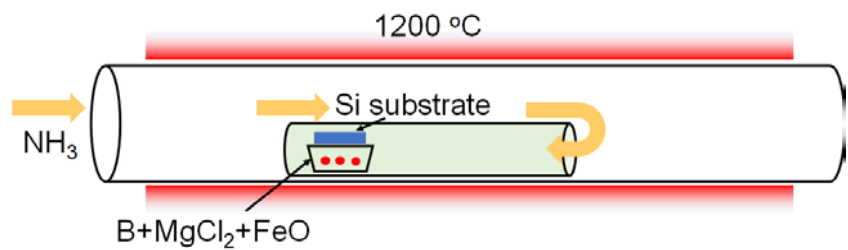

**Supplementary Figure 8 | Experimental setup for the CVD growth of BNNTs.**

Supplementary Figure 9 shows the scanning electron microscopy (SEM) image of the as-grown BNNTs on Si substrate, where clean and long BNNTs can be clearly observed, and the length is estimated to be longer than 10 μm. Transmission electron microscopy (TEM) indicates that these BNNTs exhibit a high-order tubular structure with uniform inner diameter along the axial direction (Supplementary Figure 9c). High-resolution TEM (Supplementary Figure 9d) demonstrates that the as-grown BNNTs are well crystallized almost without amorphous BN coatings on the surface of the sidewalls. Consist with previous results, the interlayer spacing of the BNNT walls is estimated as ~0.35 nm, suggesting the BNNT tends to crystalline with the (0001) faces along the radical direction. According to the distribution of inner diameter of BNNTs, we can confirm that these BNNTs have an average inner diameter of 5 nm (Supplementary Figure 9b).

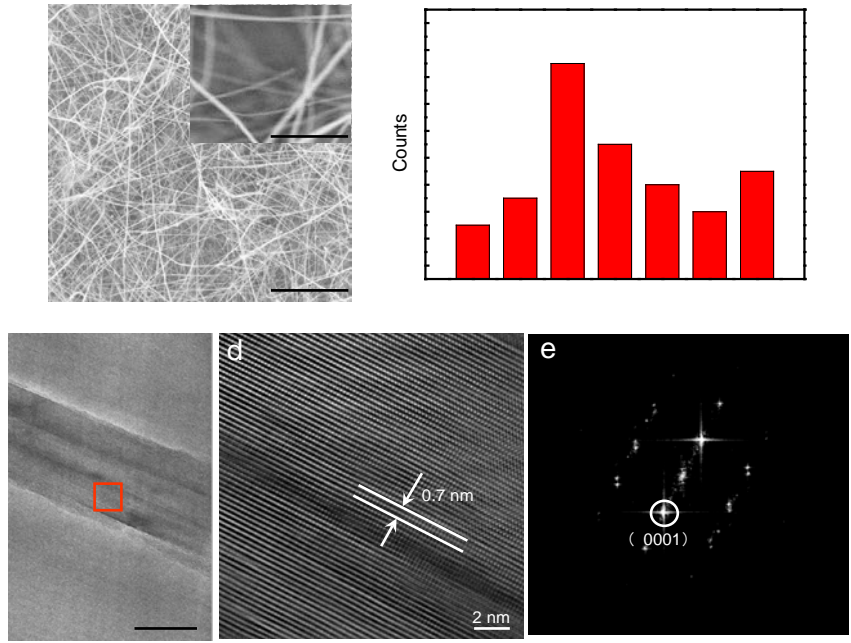

**Supplementary Figure 9 | Materials characterization of BNNTs. a,** SEM image of as-grown BNNTs on Si substrate. Inset image shows the enlarged SEM image. **b,** Statistics of the inner diameter of BNNTs. **c,** TEM of an individual BNNT showing nano-tubular structures of BNNTs with amorphous-free sidewalls. **d,** HRTEM of a BNNT with the interlayer distance to be 0.35 nm. **e,** Corresponding SEAD pattern of BNNT.

Chemical etching combined with thermal treatment was first utilized to open the ends of BNNTs, which ensures an efficient filling of Te NWs in the cavity of BNNTs. However, due to the strong interactions at Te/BNNT interface raised from the delocalization of π electrons in BNNTs, the high aspect ratio filling of Te is restricted, and only 50% of the BNNTs are successfully filled with Te NWs with the filling length less than 400 nm.

Te NW confined in BNNT would crystallize homogeneously in form of a single crystal. Supplementary Figure 10a shows the TEM characterization of an individual Te-BNNT, where the Te NW can be uniformly filled into the cavity of BNNT. Supplementary Figure 10b and 10c represent the HRTEM images of Te-BNNT obtained in the areas marked with red rectangular in Supplementary Figure 10a. The lattice fringes belonged to Te NW can be clearly observed, suggesting that Te NW is well crystallized with high quality. The single-crystalline Te NW could be further confirmed by the corresponding SEAD pattern (inset image in the upright corner) collected from HTREM, where only single set of diffraction can be detected.

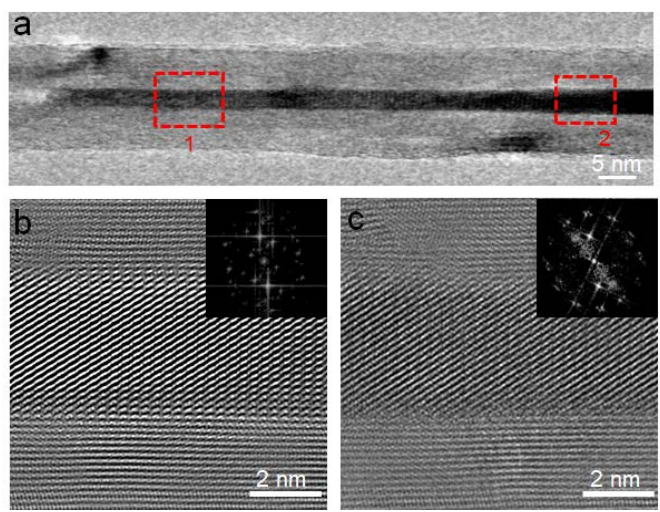

**Supplementary Figure 10 | TEM characterization of Te-BNNT. a,** low-magnificent TEM image of an individual Te NW encapsulated by BNNT. **b-c,** HRTEM images collected from the areas marked with rectangular in Supplementary Figure 10a. Inset image shows the corresponding SEAD patterns of single-crystalline Te NW.

The EDX mapping could strongly demonstrate that most of Te crystals are filled inside the BNNT instead of attached on the surface. As shown in Supplementary Figure 11a, no Te residuals could be detected on the outer surface of BNNT. In order to eliminating the influence of Te clusters, plasma treatment was first utilized to clean the outer surface of BNNT after growth. Supplementary Figure 11b shows the Raman spectrum of Te-BNNT before and after plasma cleaning, and no obvious change was detected. The $A_1$ and $E_2$ peaks of Te-BNNT exhibit a clear shift toward high frequency compared with bulk Te, which means that the signal collected is originated from coral Te NW instead of bulk Te cluster attached on BNNT surface. In addition, in the modified device fabrication process, plasma etching with high power density is required to thinning top layers of the whole length of BNNT, which would also etch away the bulk Te cluster attached on surface.

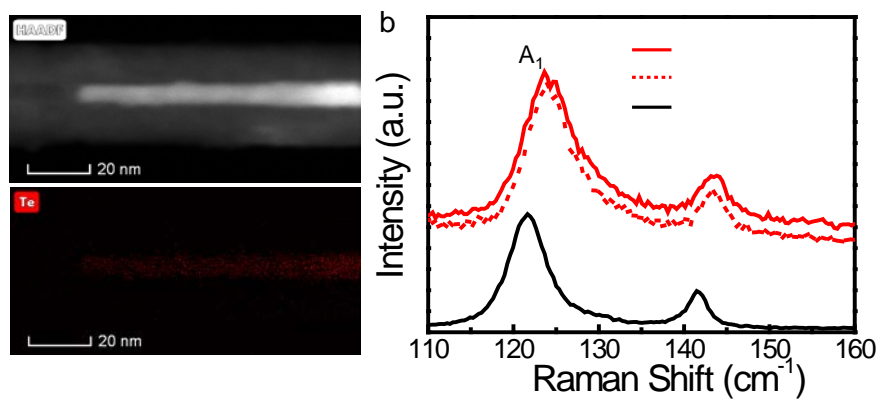

**Supplementary Figure 11 | Material characterization of Te-BNNT. a,** EDX mapping image of Te-BNNT. **b,** Comparison of Raman spectrum obtained from bulk Te, as-grown Te-BNNT and Te-BNNT after plasma cleaning.

## 4. Electrical measurements of FET devices based on bare Te NW.

In order to investigate the shielding effect of BNNTs for Te atomic chains, electrical measurements of bare Te nanowires (NWs) were also conducted for comparison. Te NWs were obtained by a solution-based process (see detailed in **Methods** section), and they need to be transferred onto a $SiO_2$/Si substrate for further characterization (Supplementary Figure 12a). The Te NWs are highly crystallized with diameter ranging from 4 nm to 50 nm as confirmed by AFM and HRTEM (Supplementary Figure 12b and 12c).

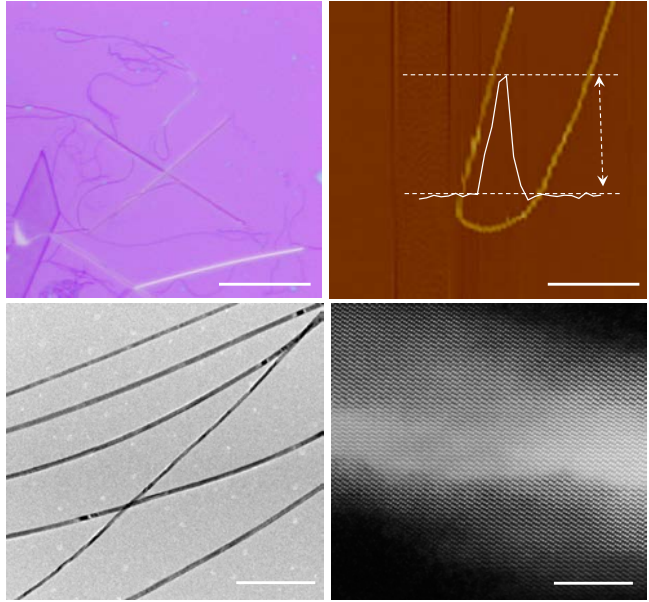

**Supplementary Figure 12 | Characterization of solution synthesized Te NWs. a,** Optical microscopy and **b,** corresponding AFM height profile of bare Te NW. **c,** Low magnitude TEM and **d,** HRTEM image of a Te NW.

Supplementary Figure 13a and 13b present the electrical performance of a 13-nm-diameter Te NW FET with a channel length of 100 nm measured at room temperature. The device exhibits the typical p-type characteristics with on/off ratio on the order of ~$1\times10^2$ due to distinct short-channel effects and narrow bandgap of Te. The drain current can reach over 900 mA/mm due to the high hole mobility of Te.

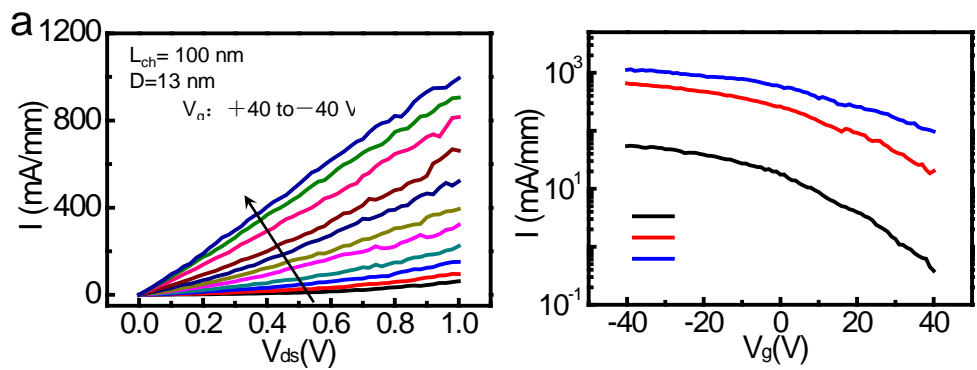

**Supplementary Figure 13 | Electrical performance of a Te NW transistor. a,** output curves and **b,** transfer curves of the device with a short channel of 100 nm.

The current capacity of solution synthesized Te NWs was studied based on 100 nm channel devices. As shown in Supplementary Figure 14, the average value for bare Te NWs is estimated be $7.6\times10^6$ A cm$^{-2}$, while the maximum ampacity value for the Te-BNNTs can be significantly improved up to $1.5\times10^8$ A cm$^{-2}$, almost two orders of magnitude larger than the bare ones.

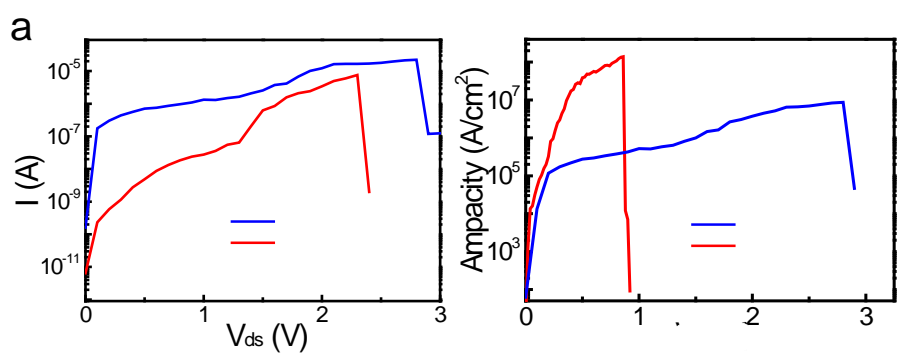

**Supplementary Figure 14 | I-V curves up to breakdown of devices. a,** Te NW devices with different diameters. **b,** Te NW and Te-BNNT devices with the same diameter.

The diameter-dependence on/off ratio and maximum drain current based on long-channel Te NW devices are presented in Supplementary Figure 15, where the data are elucidated from more than thirty bare Te NW devices. Due to the enhanced gate electrostatic control in narrow nanowires, the on/off ratio increased sharply from ~$1\times10^2$ to ~$1\times10^5$ as the diameter decreases from 40 to 6 nm. However, the narrower samples exhibit the opposite tendency with smaller current density since they are more susceptible to surface oxidation and electrical defects. Indeed, FETs based on ultra-narrow Te NWs (less than 6 nm) cannot sustain electrical conduction and gate modulation since the crystalline structure of Te NW cannot be preserved well during fabrication and measurements in ambient.

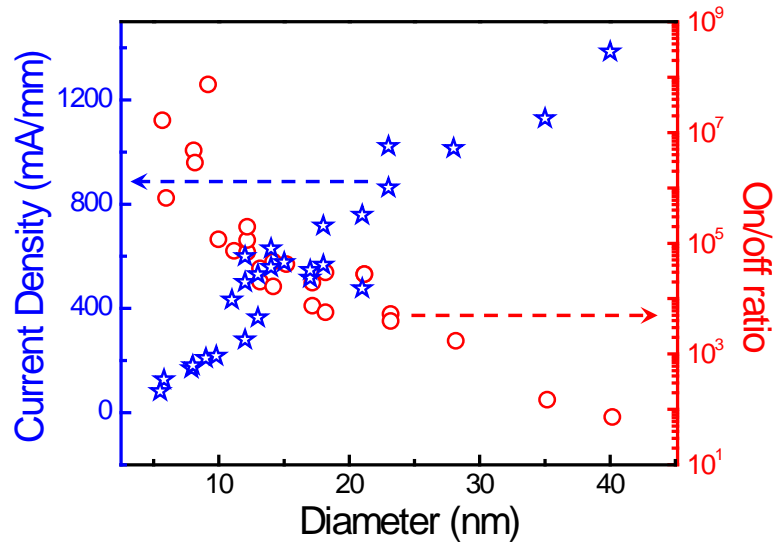

**Supplementary Figure 15 | Diameter-dependent electrical performance of Te NW transistors.**

## 5. Thermal stability and electrical properties of Te-BNNTs.

*In-situ* temperature-dependent Raman spectroscopy was conducted to investigate the thermal stability of Te-BNNTs and bare Te NWs. The Raman signals were collected after the samples annealed for 5 min in ambient environment. As shown in Supplementary Figure 16, bare Te NWs are easily degraded and damaged at 250 ºC, which is much lower than that for those shielded by BNNTs (400 ºC). The shift of Raman peaks can be attributed to the non-crystalizing phase change induced by high temperatures.

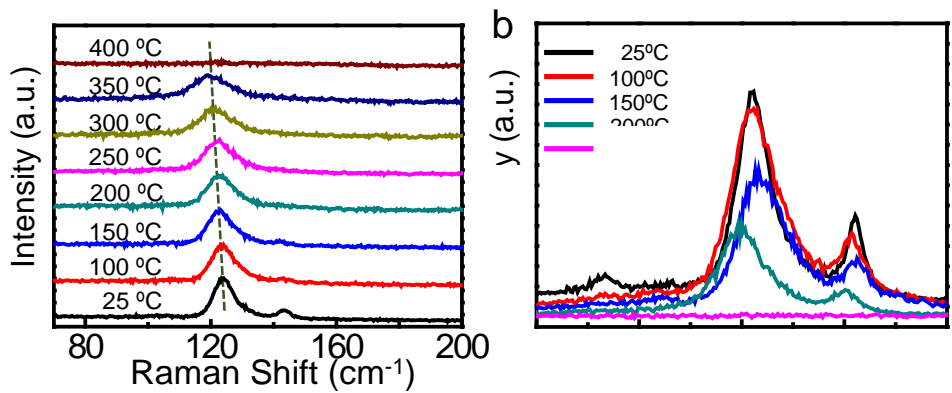

**Supplementary Figure 16 |** *In-situ* **temperature-dependent Raman spectra. a,** Te-BNNTs and **b,** bare Te

NWs.

Plasma etching of the BN shielding layer is very sensitive to the etching time. As shown in Supplementary Figure 17, after 30 s of etching, the thickness of BNNT decreases from 27 nm to 16 nm, and the top BN layer could be completely removed, resulting in a Ni/Te contact at interface. The over etching of BN deteriorates the performance of Te-BNNT devices and leads to a smaller breakdown current of $2.3\times10^{-6}$ A, compared with $17.4\times10^{-6}$ A obtained from the device with optimized 20 s etching time.

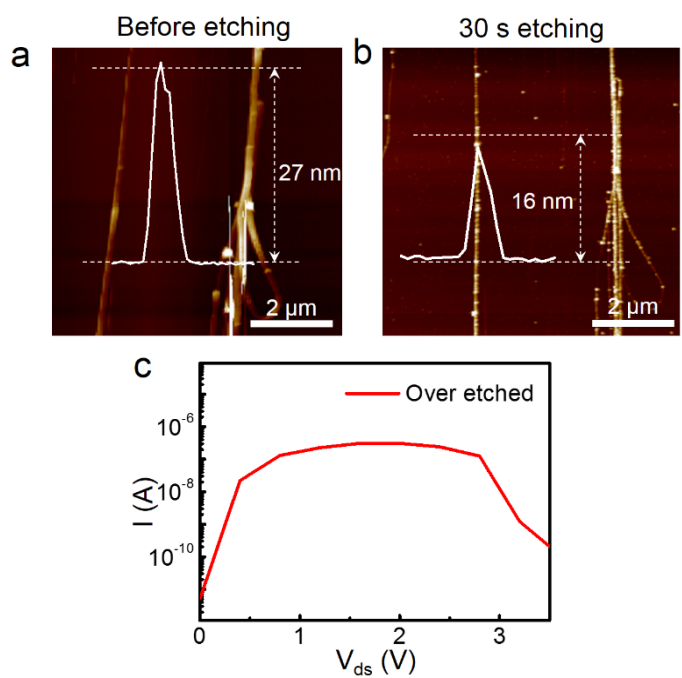

**Supplementary Figure 17 | Electrical performance of Te-BNNT devices after plasma etching.** **a-b**, AFM height profile of Te-BNNT before and after 30 seconds plasma etching. **c,** I-V curve of a Te-BNNT device after 30 seconds of etching.

The current capacity of semiconducting 1D nanowires were summarized in Supplementary Table 1. This value of $1.5 \times 10^8$ A/cm$^2$ for Te-BNNT exceeds most of the semiconductor NWs, and it is even higher than the most-studied quasi-1D metallic nanowires with high-ampacity (such as TaSe$_3$, ZrTe$_3$). Indeed, it is only slightly smaller than the semiconducting SWCNTs with the highest value of $3\sim9 \times 10^8$ A/cm$^2$.

**Supplementary Table 1 | Comparison of ampacity of devices based on semiconducting nanowires**

| Semiconductor | $L_{ch}$ (μm) | D (nm) | $I_{max}$ (μA) | Ampacity (MA/cm$^2$) |
|---|---|---|---|---|
| Si | 5.06 | 180 | 170 | 0.1 |
| PbSe | 3 | 80 | 10 | 0.5 |
| Ge | 3 | 20 | 0.7 | 0.56 |
| SnO$_2$ | 2.2 | 50 | 25 | 1 |
| GaN | 3.2 | 400 | 3200 | 3 |
| InP | 2 | 100 | 1000 | 10 |
| InAs | 1.5 | 33 | 96 | 10 |
| NiGe$_2$ | 1.6 | 50 | 850 | 35 |
| ZnO | 3 | 228 | 1000 | 63 |
| SWCNT | 0.02 | 1.2 | 10 | 880 |
| SWCNT | 0.3 | 1.2 | 3~6 | 264~528 |
| **Te-BNNT** | **0.1** | **5** | **29** | **150** |

The BNNT shielding of Te NWs enables the FETs to functionalize even with the Te diameter of 2 nm. As shown in Supplementary Figure 18, the FET exhibits p-type transport behaviors, and the on-state current could reach up to 35 mA/mm at $V_{ds}$=4V. The on/off ratio at drain bias of 1V is over $1\times10^2$, which is still a decent value considering the short-channel effect, narrow bandgap of ~0.4 eV and Schottky-like large contact resistance. Similar to n-type Te-BNNT FETs, $I_{ds}$ increases exponentially at higher $V_{ds}$, showing a typical Schottky barrier like characteristics, due to the inserted thin BN layer between Ni and Te.

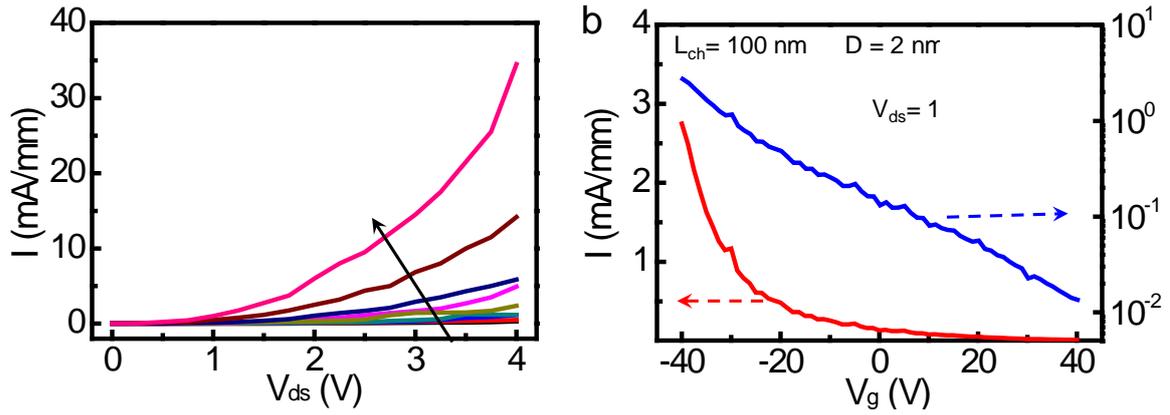

**Supplementary Figure 18 | Electrical characteristics of a p-type Te-BNNT transistor with a channel length of 100 nm. a,** output curves of the device. **b,** corresponding transfer curves of the same device at $V_{ds}$=1V.

## 6. Contact resistance in Te-BNNTs devices.

It is essential to evaluate the channel mobility ($\mu_{ch}$) of Te atomic chains. However, the large contact resistance due to the tunneling resistance of BN layer and general difficult to contact NWs would underestimate the intrinsic mobility ($\mu$) of Te atomic chains. Thus, we need to eliminate the influence of contact resistance ($R_c$). For FET devices, two techniques can be employed to correct μ. The transfer length measurement (TLM) requires multiple devices of varied L, thus it is only plausible for large-area samples. On the contrary, the Y function method (YFM) can be conducted based on individual devices, which is much more suitable for Te-BNNT FETs with short channel[5,6].

The $I_{ds}$ in linear region could be described as following:

$$I_{ds} = \frac{W}{L}C_{ox}\mu_e(V_{gs} - V_{th})V_{ds} = \frac{W}{L}C_{ox}\frac{\mu_0}{1 + \theta(V_{gs} - V_{th})}(V_{gs} - V_{th})V_{ds}$$

The $\mu_e$, $\mu_0$, $C_{ox}$, $V_{th}$, $W$, $L$, and $\theta$ represents the effective mobility in linear regime, the intrinsic mobility, the capacitance between the channel and the gate per unit area, the threshold voltage, the channel width, the channel length, and the mobility attenuation coefficient, respectively, and the Y-function can be defined as

$$Y - Function = \frac{I_{ds}}{\sqrt{G}} = \sqrt{\frac{WC_{ox}V_{ds}\mu_0}{L}} \cdot (V_{gs} - V_{th})$$

Where the transconductance $G$ could be described by

$$G = \frac{\partial I_{ds}}{\partial V_{gs}} = \frac{WC_{ox}V_{ds}\mu_0}{L} \frac{1}{[1 + \theta(V_{gs} - V_{th})]^2}$$

Based on the transfer curve at $V_{ds}$= 0.8 V, we can linear fit the derived quantity Y function in the sublinear regime of on state (Supplementary Figure 19). From the slope of Y-function, we can extrapolate $\mu_0$ value at $V_{ds}$=0.8V, which is independent of the attenuating factors, was 1.85 cm$^2$ V$^{-1}$s$^{-1}$. The mobility attenuation factor $\theta$ can be described as

$$\theta = \theta_{ch} + \theta_c = \theta_{ch} + \mu_0 C_{ox} R_c \frac{W}{L}$$

where $\theta_{ch}$, $\theta_c$, and $R_c$ denote the mobility attenuation factor from the channel, mobility attenuation factor from the contact and the contact resistance, respectively. Assuming that $\theta_{ch}$ is negligible, the value of $\theta$ is estimated to be 0.007. The corresponding gate-dependent resistance is plotted in Supplementary Figure 19d, $2R_c$ at high $V_g$ was estimated to be 12 KΩ · μm. This value is much higher than widely studied metal/2D systems, which is supposed to be introduced by the inserted BN layer.

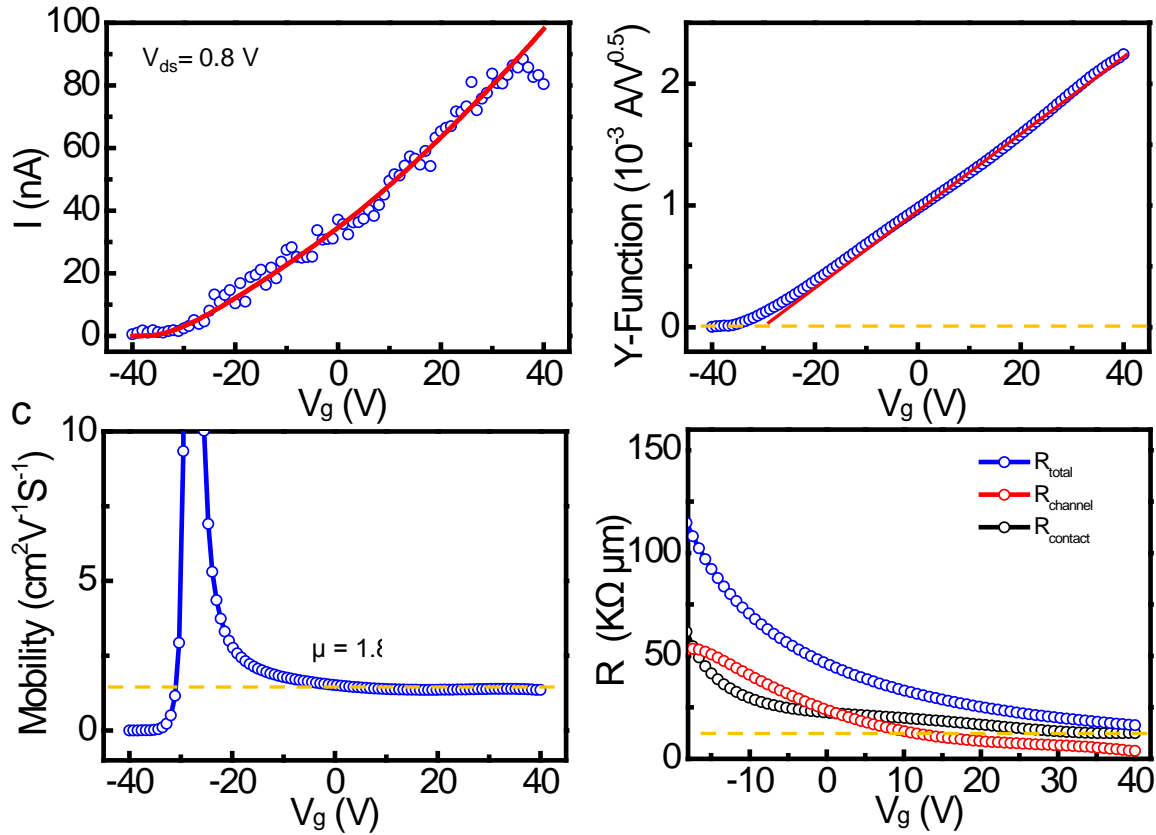

**Supplementary Figure 19 | Estimation of carrier mobility and contact resistance using Y-function method. a,** The transfer curve of a FET with 2 nm Te atomic chains as the conducting channel at $V_{ds}$=0.8 V. **b,** Y-function plots of the device. **c,** The calculated intrinsic carrier mobility of the Te atomic chains. **d,** The gate dependent resistance of the device.